\documentclass[arxiv]{medphyspaper}
\usepackage{csquotes}

\normalsize
\usepackage{booktabs}
\usepackage{array}
\newcolumntype{M}[1]{>{\centering\arraybackslash}m{#1}}
\usepackage{acro}
\acsetup{make-links=false}

\usepackage{siunitx}  
\graphicspath{{figures/}}
\DeclareGraphicsExtensions{.pdf,.jpeg,.png}
\usepackage{url}
\DeclareAcronym{BNN}{
short = BNN,
long = Bayesian neural network ,
}

\DeclareAcronym{AI}{
short = AI,
long = artificial intelligence ,
}

\DeclareAcronym{std}{
short = std,
long = standard deviation ,
}

\DeclareAcronym{BLSTM}{
short = B-LSTM,
long = Bayesian LSTM,
}

\DeclareAcronym{ML}{
short = ML,
long = machine learning ,
}

\DeclareAcronym{PET}{
short = PET,
long = positron emission tomography ,
}

\DeclareAcronym{MRI}{
short = MRI,
long = magnetic resonance imaging ,
}

\DeclareAcronym{PBA}{
short = PBA,
long = pencil beam algorithm ,
}

\DeclareAcronym{DTA}{
short = DTA,
long = distance to agreement ,
}

\DeclareAcronym{DD}{
short = DD,
long = dose difference ,
}

\DeclareAcronym{DNN}{
short = DNN,
long = deep neural network ,
}

\DeclareAcronym{IMPT}{
short = IMPT,
long = intensity modulated proton therapy ,
}

\DeclareAcronym{RSP}{
short = RSP,
long = relative stopping power ,
}

\DeclareAcronym{HU}{
short = HU ,
long = Hounsfield unit ,
}

\DeclareAcronym{RNN}{
short = RNN ,
long = recurrent neural network ,
}

\DeclareAcronym{LSTM}{
short = LSTM ,
long = long short-term memory ,
}

\DeclareAcronym{SiLU}{
short = SiLU ,
long = sigmoid linear unit ,
}

\DeclareAcronym{ReLU}{
short = ReLU ,
long = rectified linear unit ,
}

\DeclareAcronym{MC}{
short = MC ,
long = Monte Carlo ,
}

\DeclareAcronym{PB}{
short = PB ,
long = pencil beam ,
}

\DeclareAcronym{LR}{
short = LR ,
long = learning rate ,
}

\DeclareAcronym{MSE}{
short = MSE ,
long = mean squared error ,
}

\DeclareAcronym{MAE}{
short = MAE ,
long = mean absolute error ,
}

\DeclareAcronym{SMSE}{
short = SMSE ,
long = scaled mean squared error ,
}

\DeclareAcronym{KL}{
short = KL ,
long = Kullback–Leibler ,
}

\DeclareAcronym{ELBO}{
short = ELBO ,
long = evidence lower bound ,
}

\DeclareAcronym{CT}{
short = CT ,
long = computed tomography ,
}

\usepackage{todonotes}

\usepackage[capitalize]{cleveref}
\usepackage{enumitem}

\title{BayesDose: Comprehensive proton dose prediction with model uncertainty using Bayesian LSTMs}
\runningtitle{BayesDose}

\author*[DKFZ,HIRO,ICS]{Luke Voss}

\address[DKFZ]{Department of Medical Physics in Radiation Oncology, German Cancer Research Center -- DKFZ, Im Neuenheimer Feld 280, 69120 Heidelberg, Germany}
\address[HIRO]{Heidelberg Institute for Radiation Oncology -- HIRO, Im Neuenheimer Feld 280, 69120 Heidelberg, Germany}
\address[ICS]{Ruprecht Karl University of Heidelberg, Institute of Computer Science, Heidelberg, Germany}

\author*[HIRO,CCU]{Ahmad Neishabouri}

\address[CCU]{Clinical Cooperation Unit Radiation Oncology, German Cancer Research Center -- DKFZ, Heidelberg, Germany}

\author[KIT,DKFZ,HIDSS]{Tim Ortkamp}
\address[KIT]{Karlsruhe Institute of Technology (KIT), Steinbuch Centre for Computing, Karlsruhe, Germany, Heidelberg, Germany}
\address[HIDSS]{HIDSS4Health -- Helmholtz Information and Data Science School for Health, Karlsruhe/Heidelberg, Germany}

\author[HIT]{Andrea Mairani}
\address[HIT]{Heidelberg Ion‐Beam Therapy Center (HIT), Im Neuenheimer Feld 450, D‐69120, Heidelberg, Germany}

\author[DKFZ,HIRO]{Niklas Wahl}
\ead{n.wahl@dkfz.de}

\version{2}

\begin{document}
\frontmatter
\date{\today}

\maketitle

\begin{abstract}	
\textbf{Purpose:} 
Fast dose calculation techniques are needed in proton therapy, particularly in light of time restrictions in adaptive workflows. Neural network models show the potential to substitute conventional dose calculation algorithms with fast and accurate dose predictions, while lacking measures to quantify an individual prediction's quality. We propose to use a Bayesian approach to learn uncertainty of AI-based dose prediction.\\
\textbf{Methods:} 
Our resulting BayesDose-Framework is based on a previously published deterministic LSTM. Similarly, it is trained and evaluated on Monte Carlo beamlet doses simulated on (1) \num{2500} water phantoms with slab inserts and (2) \num{1000} geometries extracted from a lung patient for a single initial energy. The network's weights are parameterized with 2D Gaussian mixture models, and \num{100} ensemble predictions are used to quantify mean dose predictions and their standard deviation. Generalizability as well as re-training of the model is evaluated on smaller datasets with two different initial energies as well as five additional patients.\\
\textbf{Results:} 
The averaged predictions of the BayesDose model performed similarly to its deterministic variant and at least as good the original published \ac{LSTM} model. Predictions of the uncertainty (measured through the sampled predictions' standard deviation) seemed conservative, particularly for the phantom dataset, however regions of high uncertainty correlated spatially with the largest dose differences between the prediction and Monte Carlo calculated reference. Large average uncertainty within a prediction correlates strongly with dosimetric differences (up to $\rho = \num{-0.74})$. This correlation is reduced when applying the model to patient data with unseen HU value ranges. Runtime overhead could be decreased to 9x of a deterministic prediction for an ensemble size of \num{100} by parallelizing predictions and presampling network weights.\\
\textbf{Conclusion:}
Bayesian models for dose prediction can produce fast predictions with quality equal to analogue deterministic models. The obtained prediction's standard deviation correlates well -- both globally and locally -- with dosimetric inaccuracy. Models like BayesDose could thus support decision making and quality assurance when translating dose prediction models into the clinic, while the Bayesian approach in general could translate to other AI models in medical physics.

\end{abstract}


\mainmatter

\section{Introduction}
Dose calculation for particle therapy is a computational task sensitive to runtime and accuracy. Algorithms for dose calculation usually present a trade-off between these, with \ac{MC} dose calculation on the one side as the gold standard for accuracy but deficits in runtime, and pencil-beam dose calculation on the other side as one of the fastest, but also less accurate methods. The latter are still used in the context of treatment plan optimization regularly, requiring the computation of each potential beamlet's dose. While one may argue that \ac{MC} dose calculation is reaching runtimes sufficient for inverse planning, treatment planning is at the same time evolving towards even more demanding time restrictions, e.\,g., within real-time/online adaptation, tightening the \enquote{need for speed}.

To combine accuracy with short runtimes, recent research proposed the use of \ac{AI} to either correct the result of fast, inaccurate algorithms\cite{Xing2020,Wu2021} or learn the dose calculation entirely \cite{Neishabouri2021,Pastor-Serrano2022,Kontaxis2020,Martinot2021}.
These works usually train on a reference dataset computed with \ac{MC} algorithms for high accuracy. This initial training step is not particularly time sensitive, allowing to build specialized models (e.\,g., per commissioned energy, patient class etc.). The inference step during the actual dose calculation task can then be implemented highly performant, ideally surpassing the fastest classical numerical methods while keeping near \ac{MC} dose calculation accuracy.

One concern in translating the use of such models into clinical application is \enquote{explainability}, including the quantification of the model's prediction accuracy.\cite{Barragan-Montero2022}
While the systematic error of, for example, a pencil-beam algorithm is explainable through the assumptions and approximations made, this is not straightforward for a neural network. 

To quantify the uncertainty of a model's prediction, Bayesian approaches can be used: In a \ac{BNN}, weights and biases (i.\,e., the model's free parameters) are stochasticized by parametric probability distributions whose shape is learned in the training process.\cite{MacKay1992,Gal2016a} After training, samples can be taken from those distributions to obtain a set of neural networks with different parameterizations. These then generate multiple dose predictions on a single input, consequently allowing the calculation of statistical information on their predicitive performance. For dose prediction with \acp{BNN}, this would result in multiple dose predictions on a given patient's \ac{CT} image, from which statistical moments like the expected dose prediction and respective standard deviation can be estimated.

Based on previous work by \citet{Neishabouri2021}, which utilized \acp{LSTM} for the calculation of dose of individual proton beamlets, we demonstrate the feasibility of a \ac{BLSTM} to generate such statistical dose predictions. Our model -- \enquote{BayesDose} --  provides a meaningful way to mitigate \enquote{black-box} concerns for clinical training and potentially helps quality assurance and decision making when using \ac{AI} based dose calculation.

\section{Materials and Methods}
Since the overall viability of using \ac{LSTM} networks in proton dose predictions has already been shown in \citet{Neishabouri2021}, our focus lies primarily on analyzing the feasibility of the Bayesian variant architecture for dose calculation tasks.

\subsection{Datasets}\label{datasets}
Our BayesDose model builds upon the collection of datasets already procured for \citet{Neishabouri2021}. Choosing the same datasets enables verification of the BayesDose model with its deterministic predecessor under the same boundary conditions. The datasets are briefly explained in this section -- a more detailed explanation of data procurement can be found in \citet{Neishabouri2021}.

The input data is always given as a three-dimensional, voxelized image containing \ac{RSP} values. The ground truth contains corresponding proton beamlet dose distributions as the desired output, simulated with \ac{MC}  using the TOPAS (Tool for particle simulations) wrapper for Geant4 \cite{Perl2012}. 

\begin{figure*}[htpb]
\centering
\includegraphics[width=\textwidth]{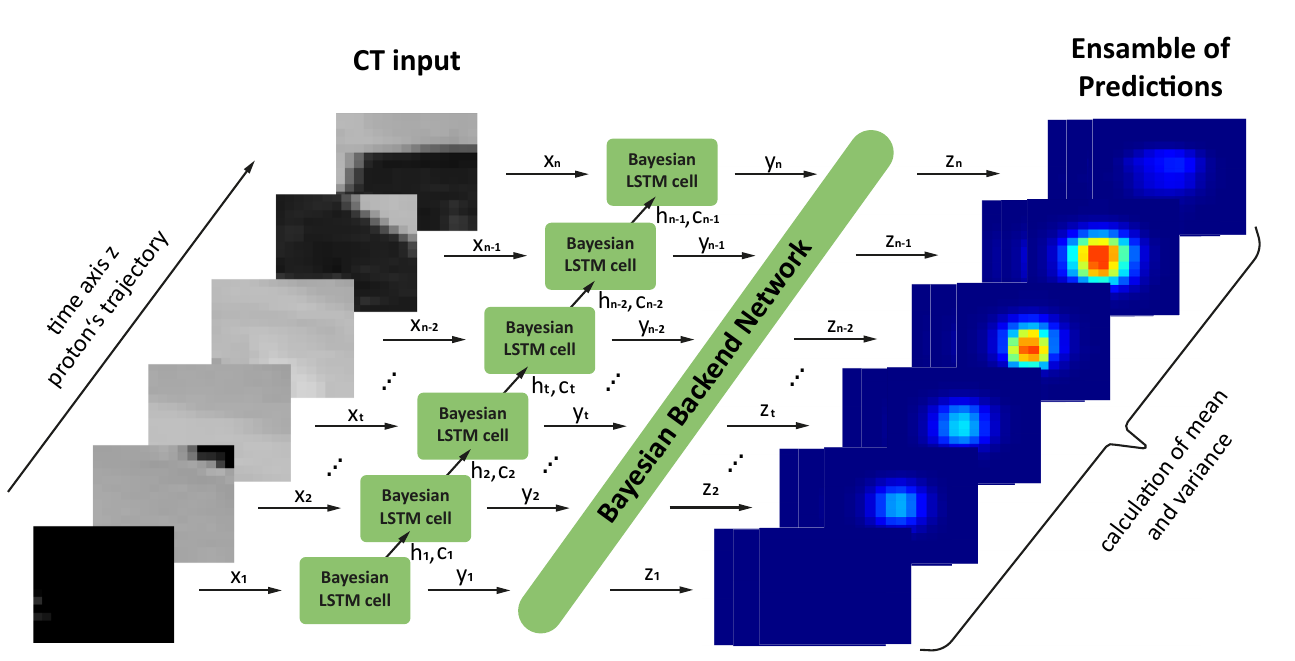}
\caption[Data processing within BayesDose]{Representation of the processing of data within the BayesDose model. The depth dimension $z$ is interpreted as the \ac{BLSTM}'s time dimension $t$: thus, each $15 \times 15$ slice of the input sequence with size $n$ is flattened into an input vector $x_t$, and then passed to the \ac{BLSTM} cell, producing a cell state $c_t$, a hidden inner state $h_t$ and output $y_t$. The output is forwarded into a fully connected Bayesian back-end network to produce the final network output $z_t$. As the network samples from its weights' and biases' probability distributions the corresponding output samples will be generated to compute, e.\,g., their mean and standard deviation. Sequence slices are taken from \citet{Neishabouri2021}.}
\label{fig:BayesianLSTM_net}
\end{figure*}

Both input and ground truth are clipped laterally and in depth to contain only the area of interest, i.\,e., the respective beamlet's dose, and is selected and rotated such that the central beamlet axis coincides with the longest dimension of the respective rectangular cuboid (compare to Figure \ref{fig:BayesianLSTM_net}). This also means that dose is always deposited from \enquote{left to right} in all figures within this work.

Overall, a supervised regression problem is defined, mapping the input \ac{RSP} values between 0 and 2.5 (0 for vacuum, 2.5 for denser bone structure) to real-valued dose output data. 

\subsubsection{Water Phantom}
\label{data:phantom}
The first dataset within the collection of \citet{Neishabouri2021} is based on dose simulations for protons with initial energy of \SI{104.25}{\mega\electronvolt} within an artificial cubic water box phantom. Inside this phantom, cuboid inhomogeneities of varying dimensions (2 to \SI{14}{\milli\m} in z’ and x’ axis) and densities (\num{0.1} \ac{RSP} to \num{2.5} \ac{RSP}) were placed. 
This dataset enables intuitive and explainable feasibility checks for developed models, and is inspired by previous dose calculation studies investigating dose calculation accuracy \citep[e.\,g.,][]{Schaffner1999}.
For this first dataset, a total of \num{10000} samples (\num{2500} phantom geometries, each in four augmentations + corresponding Monte Carlo dose) were generated with a clipping area of $80 \times 15 \times 15$ voxel with an isotropic resolution of $(\SI{2}{\mm})^3$.

\subsubsection{Lung Patient}
\label{data:lung}
The second dataset from \citet{Neishabouri2021} is generated on a lung patient case (with similar initial energy of \SI{104.25}{\mega\electronvolt}), naturally exhibiting strong anatomical heterogeneity between normal tissue, lung tissue, and bony anatomy.  Dose calculation in lung is known to be among the most difficult to achieve high accuracy with approximate methods \cite{Taylor2017}.

All data points inside this dataset are from the same patient. Different geometric problems were created by altering the beam orientation in \ang{5} steps from \ang{0} to \ang{355} as well as shifting the isocenter position in \SI{10}{\milli\m} shifts, spanning the lung along the z’ axis. This way of generating the training data resulted in cases where the gantry angle is oblique in relation to the CT axis. Consequently, these cubes experience strong wavering behavior due to the occurring interpolation errors (example can be seen in figure \ref{fig:worst_patient}). However, we decided to include these samples with high interpolation artifacts for validity of our comparison to the deterministic LSTM and to analyze how these interpolation artifacts are interfering with the model uncertainty predicted by the BayesDose model.

In total \num{4000} different samples were generated with a longitudinal clipping of $l=150$ voxels, thus creating input and output dimensions of $150 \times 15 \times 15$ at a voxel spacing of \SI{2}{\milli\meter}. 

\subsubsection{Data for generalization tests}
\label{data:generalization}
Choosing a distinct energy for training is justified by the ability to train multiple models on different initial energies (i.\,e., the commissioned energy spectra for the respective accelerator). To underline this transferability to other energies, the dataset collection includes smaller datasets for a low-range energy of \SI{67.85}{\mega\electronvolt} and a high-range energy of \SI{134.68}{\mega\electronvolt}. Each energy is represented by a total of \num{1000} pencil beam samples that were split and preprocessed with an identical approach to the primary lung patient dataset. The high-range dataset has an extended longitudinal clipping, and therefore an increased sequence length for the \ac{BLSTM}, of $l=200$.

\begin{figure}[htpb]
\centering
\includegraphics[width=0.45\textwidth]{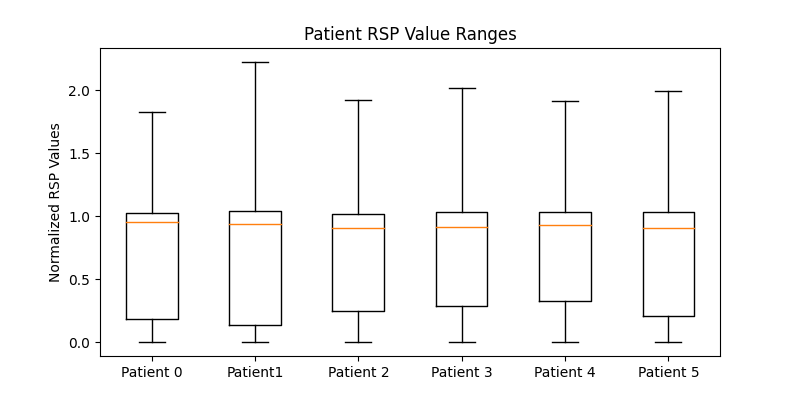}
\caption{Box plot showing the different ranges in \ac{RSP} values of the different tested patient geometries. Patient 0 denotes the original lung patient described in \cref{data:lung}.}
\label{fig:box_plot}
\end{figure}

To assess the generalizability of the model to other lung patients, the final dataset of the collection incorporates dose simulations on four additional lung patients. Each patient entails a random selection of \num{200} proton beamlets. The input data exhibits individual \ac{RSP} value ranges different to the original patient as visualized in Figure \ref{fig:box_plot}. Patients 2, 3, and 4 feature \ac{RSP} values close to the training set, patient 5 has a slightly wider \ac{RSP} range, and patient 1 has the largest range of \ac{RSP} values.

\subsection{Model Implementation}
The BayesDose model is implemented in Python 3.10.5 using the PyTorch framework 1.12.0, based on the original deterministic model from \citet{Neishabouri2021}. Bayesian network layers were implemented using the Blitz framework 0.2.8 \cite{Esposito2020}. Minimal improvements to the original reference were implemented, streamlining data processing and enabling training with larger batch sizes, providing significantly faster training time as well as improved convergence. 

\begin{figure}[htpb]
\centering
\includegraphics[width=0.45\textwidth]{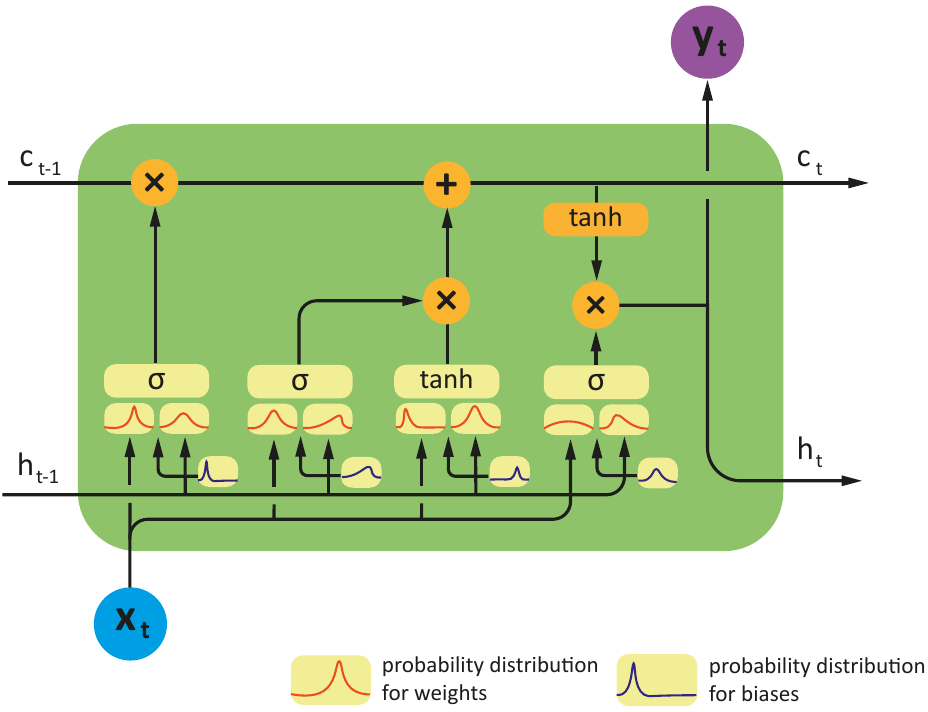}
\caption[Illustration of Bayesian \ac{LSTM} cell]{Bayesian \ac{LSTM} cell with probability distributions for weights and biases. $x_t$: input, $h_t$: hidden state (which also forms the output $y_t$), and $c_t$: cell state.
The type of activation function is indicated as sigmoid ($\sigma$) and hyperbolic tangent ($\tanh$).}
\label{fig:bayLSTM}
\end{figure}

Before outlining the full network architecture, we shortly introduce the inner workings of a \ac{BLSTM} layer (\texttt{BayesianLSTM} within the Blitz framework) compared to a conventional \texttt{LSTM} layer:
Within a \texttt{BayesianLSTM}, each weight and bias of the conventional \texttt{LSTM} layer is stochasticized by a corresponding parameterized probability distribution. The consequential Bayesian \ac{LSTM} cell structure is displayed in \cref{fig:bayLSTM}. As a result, one \ac{LSTM} cell, depending on the input and the hidden state dimension, will contain multiple stochastic weights and biases, constituting the stochastic component of the cell.

The weight update uses Bayesian inference, where a prior distribution $p(w)$ is translated into a posterior distribution $p(w|D)$ conditioned on the training data $D$ as per
\begin{equation} \label{eq:bay_inference}
p(w|D)= \frac{p(D|w)p(w)}{p(D)} \ .
\end{equation}
$p(D|w)$ is the likelihood of the weights $w$ given the data $D$, and $p(D)$ describes the evidence. Since calculating the evidence usually represents an intractable problem, the Blitz Framework approximates the posterior distribution by a variational posterior distribution parameterized with the learnable parameters $\mu$ and $\rho$. These parameters are optimized during training by minimizing the negative \ac{ELBO} defined as
\begin{equation} \label{eq:neg_elbo}
ELBO = -\underbrace{\mathbb{E}_{q_{\psi}(w)}\{\log p(D|w)\}}_\text{model fit}+\underbrace{KL\{q_{\psi}(w)||p(w)\}}_\text{regularization}\ ,
\end{equation}
which is equivalent to minimizing the \ac{KL} divergence between the variational posterior and the exact posterior distribution.
The first term $\mathbb{E}_{q_{\psi}(w)}\{\log p(D|w)\}$ measures how well the model fits the data $D$ and is determined in the implementation by the \ac{MSE}. The second term $KL\{q_{\psi}(w)||p(w)\}$ measures the distance between the variational posterior $q_{\psi}(w)$ and the prior distribution $p(w)$, acting as a regularizer, and thus preventing overfitting of the model. 

The parameters describing the probability distributions of the weights and biasas of the network are updated during training following the \enquote{Bayes by Backprob} method.\cite{Blundell2015} 

Afterwards, to obtain an ensemble of predictions, the weights and biases (indexed by $i$) of the \ac{BLSTM} network need to be sampled regarding 

\begin{equation} \label{eq:weights_sampling}
\begin{split}
w_{i}&= \mathcal{N}(0,1) \times \log (1+\rho_{i})+\mu_{i}\\
b_{i}&= \mathcal{N}(0,1) \times \log (1+\rho_{i})+\mu_{i}\ ,
\end{split}
\end{equation}

whereby $\mathcal{N}(0,1)$ represents a sampled value out of a normal Gaussian distribution with zero mean and unit variance.
The Blitz Framework uses a Gaussian prior scale mixture \cite{Blundell2015} for $p(w)$.

The prior distribution as well as the initial values of the trainable parameters $\rho$ and $\mu$ can be obtained by a hyperparameter search. To do so, we used the Optuna Framework \cite{Akiba2019} with \num{200} trials, yielding the dispersion parameters $\sigma_1 = 3.8$, $\sigma_2=0.2$, $\pi = 0.25$, $\mu=0$, and $\rho=-5.6$.

\subsection{Model Architecture}
The network architecture is predominantly consistent with the architecture of the deterministic model explained by \citet{Neishabouri2021}. For processing the input data, each $15 \times 15$ slice out of a given sequence of lateral 2D dose slices is transposed into a vector of size \num{225} and then introduced into the \ac{BLSTM} cell. Cell state and hidden state each comprise \num{1000} neurons, which, after the current slice's state update, pass on their state as input with the respective next slices. This way, each image within the input sequence gets processed inside the \ac{BLSTM} based on previously gained information. 

The final output of the \ac{BLSTM} layer is a sequence of vectors of size \num{1000}, each representing a processed image out of the input sequence. Finally, back-end fully connected layers will generate the results by converting the hidden state embedded vectors of size \num{1000} into their
original shape of size \num{225}, which after that is reshaped into a $15\times15$ slice for each $t$. 

For BayesDose, the back-end network differs from the network used by \citet{Neishabouri2021}. While each of the three Linear layers is replaced by a \texttt{BayesianLinear} layer, we also substituted the \ac{ReLU} activation functions with \ac{SiLU}\cite{Elfwing2018} to ensure smoothness of the posterior probability distributions of the weights and biases. 

The resulting back-end network has a Bayesian linear input layer of \num{1000} neurons, a hidden one of \num{100} neurons, and a Bayesian output layer of \num{225} neurons, with a \ac{SiLU} activation function between the layers.

For generating the network's final prediction and estimating the uncertainty, BayesDose generates a certain number of sampled weights and biases based on \cref{eq:weights_sampling} for the same input sequence. The resulting series of predictions can then be used to calculate the mean and the standard deviation representing the aggregate prediction of the network as well as its uncertainty estimate.

\subsection{Model Training}
The loss function used for training BayesDose is the negative \ac{ELBO} loss function composed of the \ac{MSE} and the \ac{KL} divergence (see \cref{eq:neg_elbo}). To account for the Bayesian nature, the average \ac{ELBO} loss from three calculations is used for backpropagation.

The \ac{MSE} is about a factor of $1 \cdot 10^5$ times smaller than the \ac{KL} divergence loss. This leads to the \ac{KL} divergence dominating the loss function. The optimizer then disregards minimization of the \ac{MSE} loss and thus converges slowly. To guarantee an equal impact on the loss function during optimization, we scaled the \ac{MSE} during training by above factor, and denote this scaled \ac{MSE} as \ac{SMSE}.

Performing a learning rate range test proposed by \citet{Smith2017} on the \ac{SMSE} and \ac{KL} divergence loss separately, also showed that both loss function components required a different range of optimal learning rates. To schedule learning rates appropriately, as well as benefit from its fast and low convergence capabilities, PyTorch's \texttt{OneCycleLR}\cite{Smith2019} is used.

For the phantom dataset, the starting value of the \verb|OneCycleLR| scheduler was set to a value $1.3 \cdot 10^{-3}$, allowing the \ac{SMSE} loss to converge fast and the \ac{KL} divergence loss to start converging slowly. From that point, the \ac{LR} increases until a maximum \ac{LR} is achieved. The maximum value of $6.5 \cdot 10^{-3}$  is found to be the largest \ac{LR} possible, where the \ac{SMSE} loss does not explode. Sequentially, the \ac{LR} drops to a minimum value of $4.13 \cdot 10^{-5}$  during the second half of training, causing the total loss to converge. The patient case was found to allow for slightly larger \ac{LR}s with an initial value of $3.3\cdot 10^{-4}$, a maximum \ac{LR} of $1\cdot 10^{-3}$ and a minimum \ac{LR} of $3.3\cdot 10^{-6}$.
Although \citet{Smith2019} suggests training with large batch sizes, an improvement could not be observed when applied to the BayesDose architecture.

\begin{figure}
	\centering
	\includegraphics[width=0.5\textwidth]{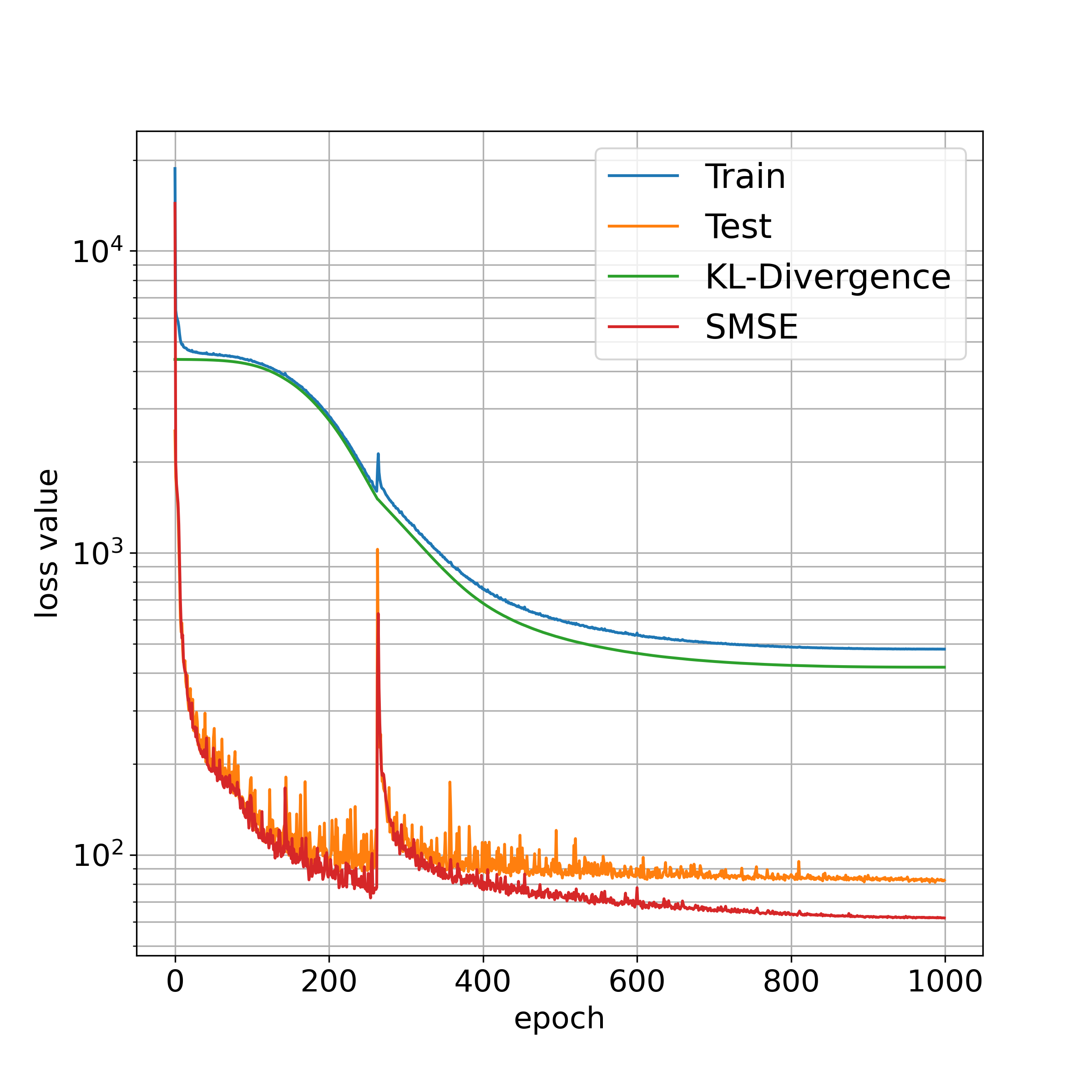}
	\caption{Loss function evolution for the \num{1000} epochs used for the patient training. The training loss is the sum of KL-divergence and \ac{SMSE}, whose contributions are also shown individually. The test loss is evaluated every 10 epochs (only \ac{SMSE}) from individual prediction samples.}
	\label{fig:loss_analysis}
\end{figure}

The learning process of the network is led by the Adam optimizer \cite{Kingma2017} with the \verb|OneCycleLR| scheduler and a batch size of \num{32}, which was found to have the smallest difference in learning rates from both loss components and ultimately offered the most effective training. The Adam optimizer is shown to generalize well across many different application area and is one of the suggested optimizers to be used for Bayesian learning \cite{Jospin2022}. The network is trained for \num{600} epochs for the phantom dataset and \num{1000} epochs for the patient dataset, after these number of epochs no significant improvements in test performance could be observed. The evolution of the total loss and its constituents is shown in \cref{fig:loss_analysis} and shows that after initial reduction of the \ac{SMSE}, the KL-divergence is gradually reduced. A full training cycle of the phantom data takes about \SI{10}{\hour} and for the patient data about \SI{11}{\hour} on an NVIDIA RTX A6000 GPU.

\subsection{Model Evaluation}
For evaluating the performance of BayesDose, \num{100} prediction samples were drawn to enable a precise estimation of the mean and standard deviation, while maintaining decent prediction time. The metrics for quantifying the performance of the model were divided into accuracy metrics using the ensemble mean and uncertainty metrics using the ensemble standard deviation. 

For the evaluation of the predictive accuracy of the network, a global $\gamma$ analysis \cite{Low1998} with a \SI{1}{\percent}/\SI{3}{\milli\meter} criterion was used. To avoid clustering of $\gamma$ pass-rates at \SI{100}{\percent}, voxels with doses under \SI{0.1}{\percent} of the maximum dose were excluded from the gamma analysis, resulting in a stricter metric than the one used in \citet{Neishabouri2021}, since values close to zero caused by numerical inaccuracy of the neural network may artificially increase the passing rate. For the patient case, 10 interpolation points were used for $\gamma$ computation. For a fair comparison, the results for the original model by \citet{Neishabouri2021} were then recalculated using this stricter criterion. In addition to the $\gamma$ pass-rate, \ac{MSE}, as well as the \ac{MAE} of the \ac{DD} are reported. 

To assess the quality of the uncertainty prediction, the relative number of voxels not correctly predicted within its respective $n\sigma$ confidence interval, where $\sigma$ is the standard deviation of the predicted dose ensemble within a voxel, was extracted for $n = [1,5]$. While the (uncertain) predicted dose is not following a Gaussian probability distribution, comparing the relative number of voxels deviating more than $n\sigma$ with the Gaussian expectation may help indicate if the model heavily over- or underestimates its confidence. Following the same data preparation technique as for the $\gamma$ analysis, \acp{DD} of less than \SI{1}{\percent} as well as differences below \SI{0.1}{\percent} of the maximum dose were neglected for the analysis.

For evaluation reference, the Bayesian \ac{LSTM} network was compared to the \ac{LSTM} network introduced by \citet{Neishabouri2021}, as well as to its equivalent deterministic structure to exclude any effects that might originate from the several differences in overall architectures (like using SiLU) as well as training beyond the switch to Bayesian Network layers.

\subsection{Experimental Design}\label{experiments}
Based on the available dataset and the implemented model, we devised five experiments to test the behavior of BayesDose. 

\begin{description}
	\item[Experiment 1:] The model is trained, tested and evaluated on the phantom dataset described in section \ref{data:phantom}. Likewise to \citet{Neishabouri2021}, a 60-20-20 split (training-validation-testing) is used.
	
	\item[Experiment 2:] The model is trained, tested and evaluated on the individual lung patient dataset described in section \ref{data:lung}, again using a 60-20-20 split.
	
	\item[Experiment 3:] Validation of the performance for different initial energies, by training, validating and testing each on the low-range energy and high-range energy of the dataset described in \cref{data:generalization}.
	
	\item[Experiment 4:] Investigation on the generalizability by applying the model trained in Experiment 2 to the other four patients. 
	
	\item[Experiment 5:] Testing the network's capability to be re-trained on new data, by first training the network on data from patient 0 and subsequently fine-tuning it for 10 epochs on patient 5, which features a wider range of \ac{RSP} values than the training dataset but still a lower range of the outlier patient 1. Afterwards, the network was tested once again on the remaining patients (similar to Experiment 4) and the results were compared.
\end{description}

In Experiments 4 and 5, we also correlate the predicted uncertainty with various $\gamma$-criteria to investigate the potential of developing decision criteria for accepting the models output or discarding it, given the output uncertainty. Experiment 5 evolves this further by testing the potential impact of re-training on previously unseen data on this correlation.

The results of this study are, if not explicitly stated otherwise, always evaluated on the unseen test subset of the data.

\section{Results}
\subsection{Phantom Data (Experiment I)}
For each individual pencil beam in the test set, the BayesDose predictions (i.\,e., the average out of 100 samples) are first compared to the \ac{MC} ground truth dose distribution using a $[1\%,\SI{3}{\milli\m}]$ $\gamma$ analysis to confirm the literature results and suitability of the architecture for dose prediction. 

In \cref{tb:error_phantom}, the average, the standard deviation, minimum and maximum of $\gamma$ pass-rates across the test dataset are reported, and the corresponding \ac{MAE} and \ac{MSE} between the generated dose cubes and the ground truth \ac{MC} simulation. 

\begin{table}[htp]
	\centering
	\caption[$\gamma$-index analysis and \acsp{MAE} and \acsp{MSE} for the phantom case]{$\gamma$-index analysis $[1\%, \SI{3}{\milli\m}]$ and \acs{MAE} and \acs{MSE} between the three trained network models and the \ac{MC} simulation on the phantom case.}
	\begin{tabular}{l c c c c}
		\toprule
		& Neishabouri & BayesDose & BayesDose \\
		& model\cite{Neishabouri2021} & (deterministic) & \\
		\midrule
		Mean $\gamma$ pass-rate [\si{\percent}] & 96.00 & 97.81 & 97.93 \\
		SD $\gamma$ pass-rate [\si{\percent}] & 2.62 & 1.90 & 1.87 \\
		Min $\gamma$ pass-rate [\si{\percent}] & 76.30 & 87.65 & 88.93 \\
		Max $\gamma$ pass-rate [\si{\percent}] & 99.27 & 100 & 100 \\
		\midrule
		\ac{MAE} [\si{\gray}] & $5.60 \times 10^{-6}$ & $2.42 \times 10^{-6}$ & $2.70 \times 10^{-6}$ \\
		\ac{MSE} [\si{\gray\squared}] & $1.02 \times 10^{-9}$ & $2.17 \times 10^{-10}$ & $2.31 \times 10^{-10}$ \\
		\bottomrule
	\end{tabular}
	\label{tb:error_phantom}
\end{table}


Both BayesDose and its deterministic variant seem well suited for dose calculation with pass-rates $>\SI{97.81}{\percent}$, even slightly outperforming the original \ac{LSTM} model by \citet{Neishabouri2021} with higher $\gamma$ average and minimum pass-rates and lower \ac{MAE} and \ac{MSE}. 
Notably, comparing the average prediction from BayesDose to the ground truth showed a small predictive performance improvement over using the deterministic variant.

\begin{figure}[htpb]
	\centering
	\includegraphics[width=0.45\textwidth]{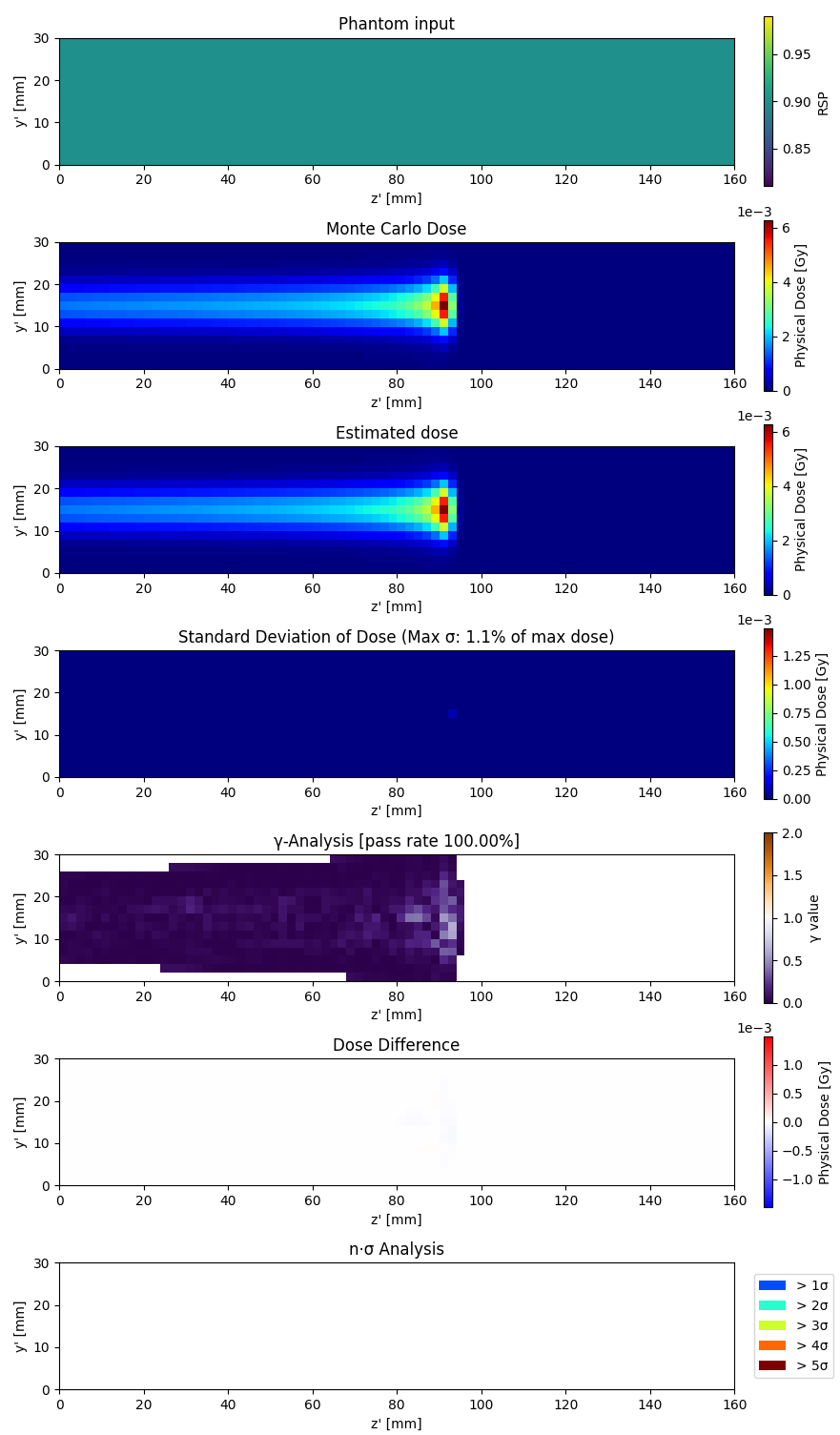}
	\caption[Best prediction of Bayesian \ac{LSTM} in the phantom case]{Prediction of the Bayesian \ac{LSTM} network with a \SI{100}{\percent} $\gamma$ pass-rate in the phantom case. The images (from top to bottom) represent the input cube, the Monte Carlo ground truth, the BayesDose prediction averaged over the ensemble, the respective standard deviation, the $\gamma$ pass-rate, the difference of Monte Carlo dose minus prediction, and the dose difference in relation to the standard deviation, where voxels deviating more than $n\sigma$ are colored accordingly. We chose this arrangement of analysis images for all subsequent displays of individual predictions.}
	\label{fig:best_phantom}
\end{figure}

Nearly perfect predictions occured when cuboid heterogeneities only had a minor effect on the dose distribution resulting in only small distortion of the Bragg-peak. Nonetheless, these samples were largely estimated by the network as cases with low output uncertainty. In the case of \cref{fig:best_phantom}, which is one of the best predictions of BayesDose, the maximum standard deviation is \SI{1.1}{\percent} (relative to the maximum dose) and no voxel differs more than $3\sigma$. 

\begin{figure}[htpb]
	\centering
	\includegraphics[width=0.45\textwidth]{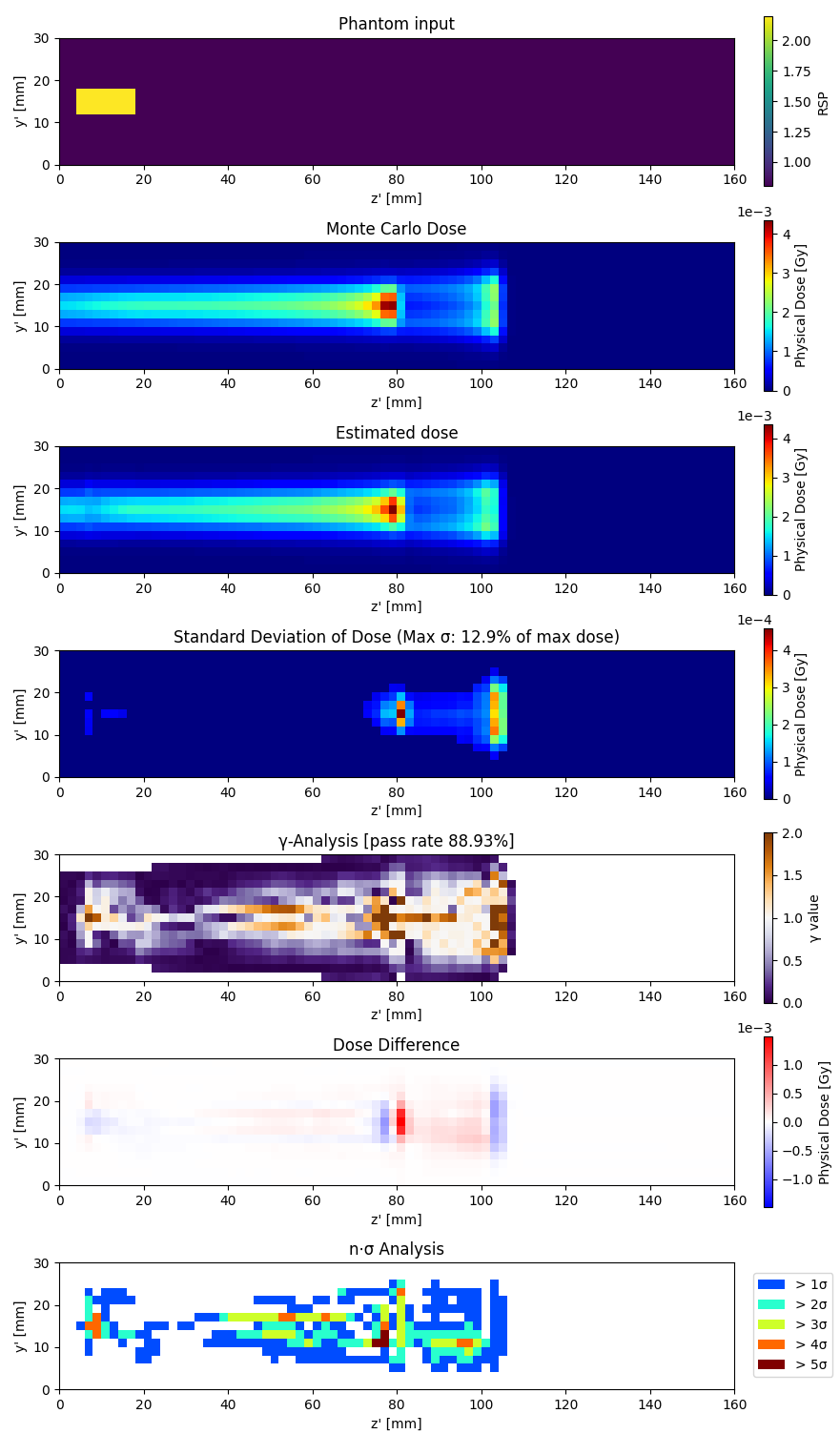}
	\caption[Worst prediction of Bayesian \ac{LSTM} in the phantom case]{Worst prediction of Bayesian \ac{LSTM} with the lowest $\gamma$ pass-rate percentage of \SI{88.93}{\percent} in the phantom case, with at the same time one of the highest standard deviations over the predictions. The figure arrangement is similar to \cref{fig:best_phantom}.}
	\label{fig:worst_phantom}
\end{figure}

The prediction with the lowest $\gamma$ pass-rate of \SI{88.93}{\percent} is shown in \cref{fig:worst_phantom}, illustrating that large dose differences between ground truth and prediction spatially coincide with high dose regions and high uncertainty. For example, at the end of the range, the standard deviation from BayesDose reached up to \SI{12.9}{\percent} relative to the maximum dose in \cref{fig:worst_phantom}, which is among the highest observed standard deviations in Experiment 1 and indicates a less robust prediction than \cref{fig:best_phantom}. \SI{0.88}{\percent} of voxels differ more than $3\sigma$ and are located mainly near high dose gradients in depth.

\begin{figure}[htpb]
	\centering
	\includegraphics[width=0.45\textwidth]{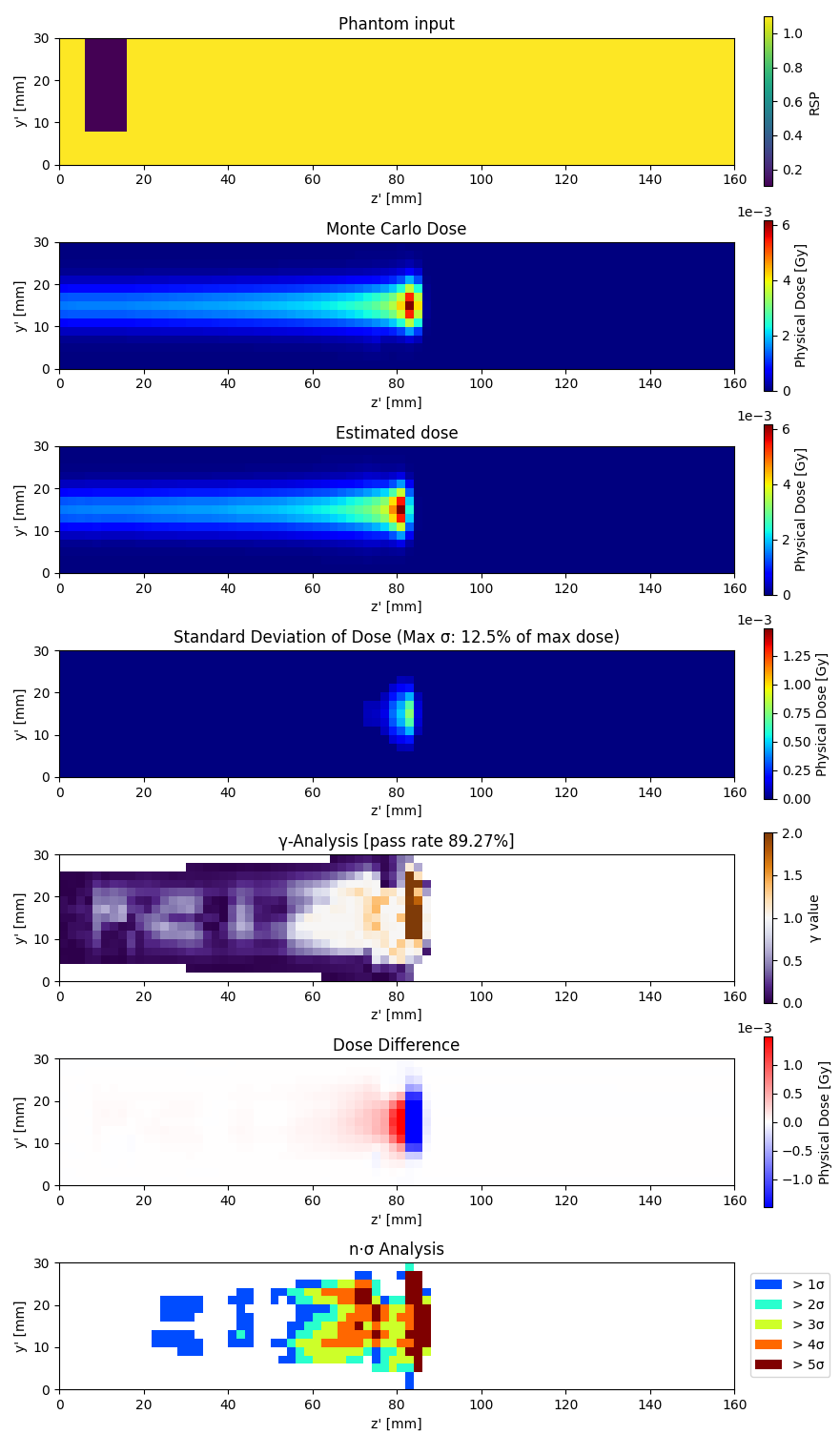}
	\caption[Worst quantification of uncertainty for the phantom case]{Worst quantification of underlying uncertainty with the largest percentage of voxels significantly deviating ($\SI{3.62}{\percent} > 3\sigma$) for the phantom case. The figure arrangement is similar to \cref{fig:best_phantom}.}
	\label{fig:worst_sigma}
\end{figure}

The obtained percentages of voxels outside of $n\sigma$ over the entire test set range are reported for each sigma from one to five in \cref{tb:sigma_phantom}. 
\begin{table}[htpb]
	\centering
	\caption[Percentage of voxels deviating more than $n\sigma$ for the phantom case]{Percentage of voxels deviating more than $n\sigma$ for the phantom case.}
	\begin{tabular}{c c c c c}
		\toprule
		\multicolumn{5}{c}{Voxels [\si{\percent}] deviating more than}\\
		$1\sigma$ & $2\sigma$ & $3\sigma$ & $4\sigma$ & $5\sigma$\\
		\midrule
		1.38 & 0.48 & 0.19 & 0.09 & 0.04 \\
		\bottomrule
	\end{tabular}
	\label{tb:sigma_phantom}
\end{table}
These values allow comparison to the Gaussian assumptions of probability mass in a confidence interval. For the phantom case, these values are consistently too low, indicating an overestimation of the uncertainty. \Cref{fig:worst_sigma} shows the case with the largest amount of voxels deviating more than $3\sigma$.

The high percentage of significantly deviating voxels in \cref{fig:worst_sigma} originates from the systematic underestimation of proton path length; the Bragg peak and therefore high dose values appear distally shifted. This indicates a failure of the model to correctly maintain the unusually long \enquote{temporal}, i.\,e., downstream, dependency on the entrance cavity in this particular case. Still, the failing area is also estimated to have high uncertainty, however, its magnitude is too small in scale to grasp the full \ac{DD} occurring in that area.

\subsection{Patient Data (Experiment 2)}
\Cref{tb:error_patient} summarizes the performance of BayesDose, its deterministic variant and the original model for the first lung patient. All three algorithms have equal performance on the test dataset with average $\gamma$ pass-rates over \SI{99.3}{\percent} and only minor differences in their \ac{MSE} and \ac{MAE} values. 
\begin{table}[htpb]
\centering
\caption[$\gamma$-index analysis and \acsp{MAE} and \acsp{MSE} for the phantom case]{$\gamma$-index analysis $[1\%, \SI{3}{\milli\m}]$ (top) and \acs{MAE} and \acs{MSE} between the three trained network models models and the \ac{MC} simulation (bottom).}
\begin{tabular}{l S S S}
\toprule
 & {Neishabouri} & {BayesDose} & {(deterministic)}\\
 $\gamma$ analysis & {LSTM\cite{Neishabouri2021}} & {\ac{BLSTM}} & \\
\midrule
Mean [\si{\percent}] & 99.60 & 99.70 & 99.59 \\
SD [\si{\percent}] & 0.62 & 0.60 & 0.78 \\
Min [\si{\percent}] & 95.89 & 95.02 & 91.44 \\
Max [\si{\percent}] & 100 & 100 & 100 \\
\midrule
\ac{MAE} [\si{\gray}] & 1.62e-5 & 1.46e-5 & 1.47e-5 \\
\ac{MSE} [\si{\gray\squared}] & 3.14e-9 & 3.70e-9 & 3.28e-9 \\
\bottomrule
\end{tabular}
\label{tb:error_patient}
\end{table}

Again, we relate dosimetric accuracy to predicted uncertainty through best and worst examples.

\begin{figure}[htpb]
	\centering
	\includegraphics[width=0.45\textwidth]{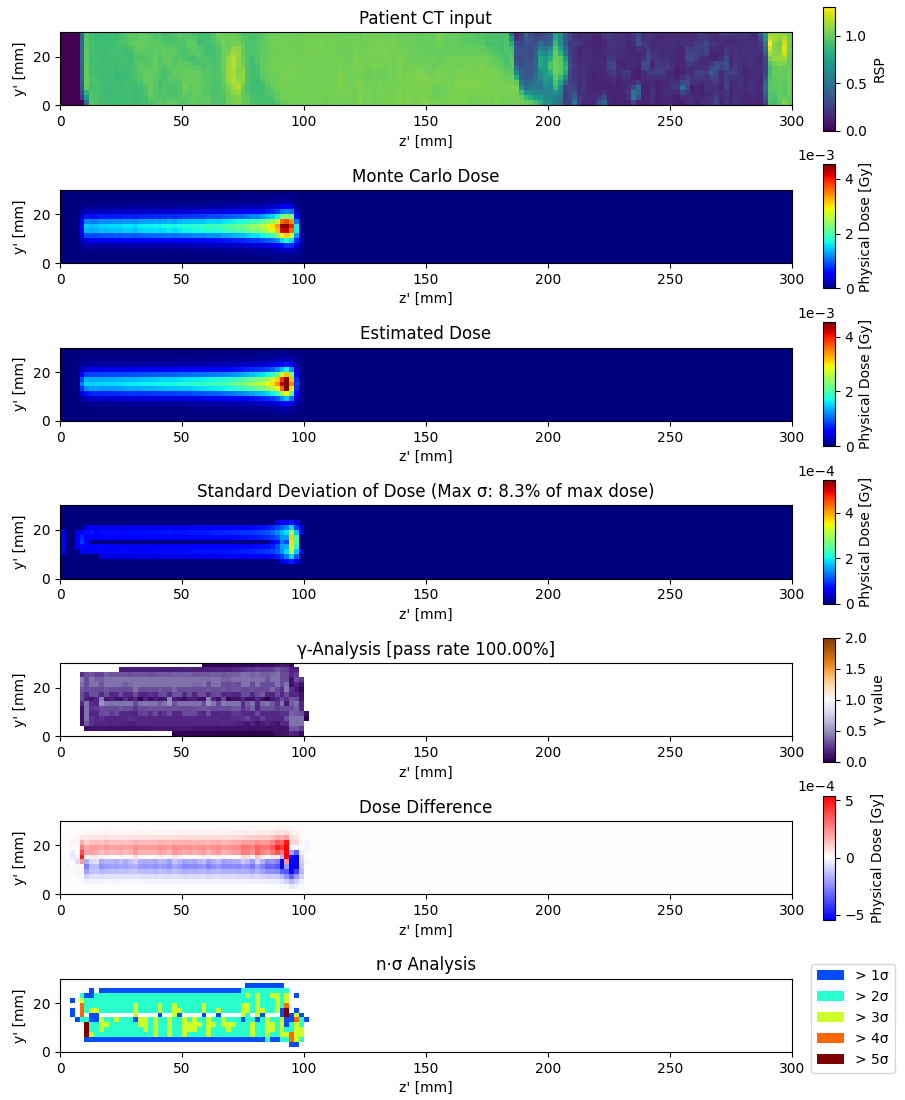}
	\caption[Best prediction of BayesDose in the patient case]{Prediction of BayesDose with a \SI{100}{\percent} $\gamma$ pass-rate of the patient dataset. The figure arrangement is similar to \cref{fig:best_phantom}.}
	\label{fig:best_patient}
\end{figure}

One of the most accurate predictions is illustrated in \cref{fig:best_patient} with a \SI{100}{\percent} $\gamma$ pass-rate. In the patient dataset, BayesDose generally predicts zero dose in low \ac{RSP} values, as the dose in air before the patient was zero in the training data as well, which reproduced the original deterministic model's behavior.

BayesDose seemingly learns the occurring interpolation artifacts from the pre-processing pipeline as model uncertainty, forming \enquote{stripes} of standard deviation around the beam line axis in where the interpolation effects were usually most pronounced in the training dataset. Also, \cref{fig:best_patient} shows a higher maximum standard deviation relative to the maximum dose (\SI{8.3}{\percent}) than the best phantom prediction (\SI{1.1}{\percent}). 

\begin{figure}[htpb]
	\centering
	\includegraphics[width=0.45\textwidth]{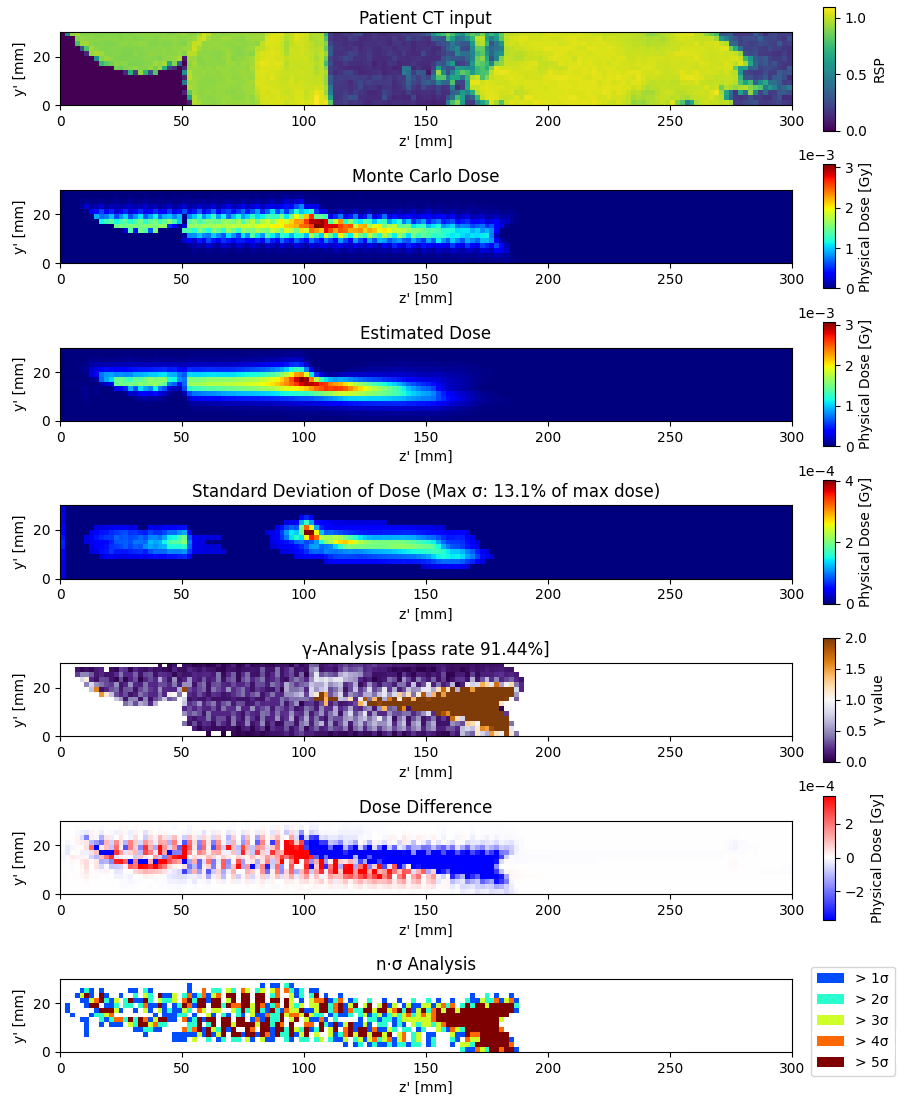}
	\caption[Worst prediction of \ac{BLSTM} in the patient case]{Worst prediction of BayesDose with the lowest $\gamma$ pass-rate percentage of \SI{88.93}{\percent} from the patient dataset. Again, the worst prediction exhibits one of the highest observed standard deviation estimates as well. The figure arrangement is similar to \cref{fig:best_phantom}.}
	\label{fig:worst_patient}
\end{figure}

In the test sample with the lowest dosimetric accuracy, shown in \cref{fig:worst_patient}, BayesDose underestimates the length of the dose wash-out beyond the Bragg peak, but associates this region with very high uncertainty (reaching up to \SI{13.1}{\percent} of the maximum dose). Further, the interpolation artifacts seem to be a large driver of failing voxel dose predictions, visible in the large number of voxels deviating more than $5\sigma$.

\begin{figure}[htp]
\centering
\includegraphics[width=0.45\textwidth]{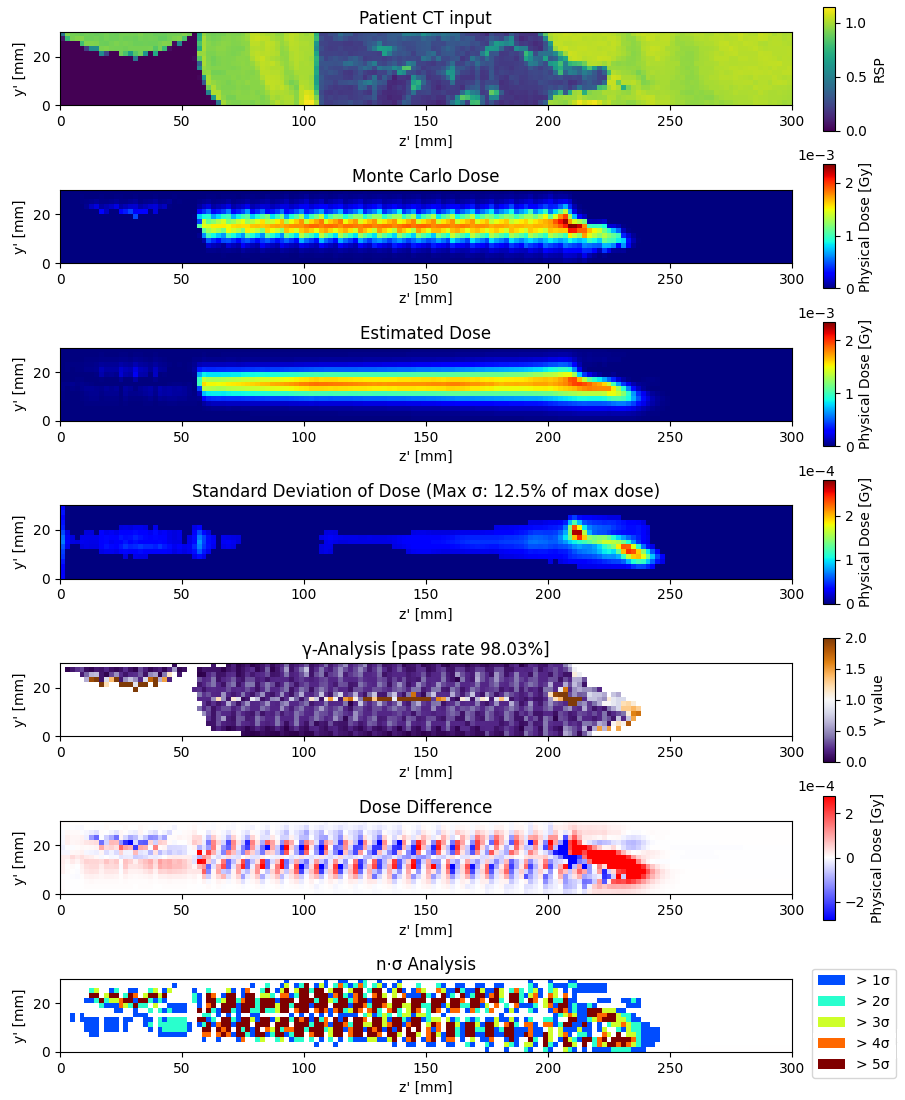}
\caption[Worst quantification of uncertainty for the patient case]{Worst quantification of underlying uncertainty with the largest percentage of voxels significantly deviating ($\SI{15.99}{\percent} > 3\sigma$) for the patient dataset. The figure arrangement is similar to \cref{fig:best_phantom}.}
\label{fig:worst_sigma_patient}
\end{figure}

In general, over the whole patient dataset,  the average percent of voxels deviating more than $n\sigma$ (shown in \cref{tb:sigma_patient}) are consistently higher  than in the first experiment on the phantom data. Thus, according to \cref{tb:sigma_patient}, BayesDose seems to produce less conservative uncertainty predictions on the patient data. A large part of voxels significantly deviating beyond $5\sigma$, as illustrated in the prediction example with the maximum amount of voxels showing a dose difference $>n\sigma$ (\cref{fig:worst_sigma_patient}), are seemingly rooted in the interpolation artifacts in the training data.

\begin{table}[htpb]
\centering
\caption[Percentage of voxels deviating more than $n\sigma$ for the phantom case]{Percentage of voxels deviating more than $n\sigma$ for the patient case.}
\begin{tabular}{c c c c c}
\toprule
\multicolumn{5}{c}{Voxels [\si{\percent}] deviating more than}\\
$1\sigma$ & $2\sigma$ & $3\sigma$ & $4\sigma$ & $5\sigma$\\
\midrule
11.45 & 7.92 & 5.56 & 3.87 & 2.67\\
\bottomrule
\end{tabular}
\label{tb:sigma_patient}
\end{table}

\subsection{Multiple Energies (Experiment 3)}
BayesDose was trained and tested on two additional proton energies. Similar to the original model by \citet{Neishabouri2021}, the accuracy is comparable across all three energies, as listed in \Cref{tb:multi_energy} where for all three energies, the mean pass-rate is above \SI{99}{\percent}. 

\begin{table}[htpb]
\centering
\caption[]{$\gamma$-index analysis on datasets with three differently energized proton beamlets ($[1\%, \SI{3}{\milli\m}]$).}
\begin{tabular}{l c c c c}
\toprule
$E$ (\si{\MeV}) & Mean(\%) & SD(\%) & Min(\%) & Max(\%)\\
\midrule
67.85 & 99.45 & 0.92 & 94.16 & 100\\
104.25 & 99.59 & 0.78 & 91.44 & 100 \\
134.68 & 99.00 & 1.13 & 92.47 & 100\\
\bottomrule
\end{tabular}
\label{tb:multi_energy}
\end{table}

\subsection{Inter-patient Generalization (Experiment 4)}
For experiment 4, that is, evaluation of the network performance on 5 additional lung patients, we again first analyze dosimetric performance of the average prediction in  \cref{tb:multi_patients}. BayesDose shows similar behavior to the original deterministic model, seeing Patient 1 and Patient 5 as worse performing outliers compared to the rest.

\begin{table}[htpb]
\centering
\caption[]{$\gamma$-index analysis on five different lung cancer patients ($[1\%, \SI{3}{\milli\m}]$). The network was trained on Patient 0.}
\begin{tabular}{l c c c c}
\toprule
 & Mean(\%) & SD(\%) & Min(\%) & Max(\%)\\
\midrule
Patient 0 & 99.59 & 0.78 & 91.44 & 100\\\midrule
Patient 1 & 96.33 & 4.04 & 73.01 & 99.98 \\
Patient 2 & 99.29 & 1.02 & 90.21 & 100\\
Patient 3 & 99.12 & 1.07 & 94.82 & 100\\
Patient 4 & 99.32 & 0.87 & 94.05 & 100\\
Patient 5 & 98.04 & 2.09 & 88.27 & 100\\
\bottomrule
\end{tabular}
\label{tb:multi_patients}
\end{table}

\begin{figure*}[htpb]
	\centering
	\includegraphics[width=\textwidth]{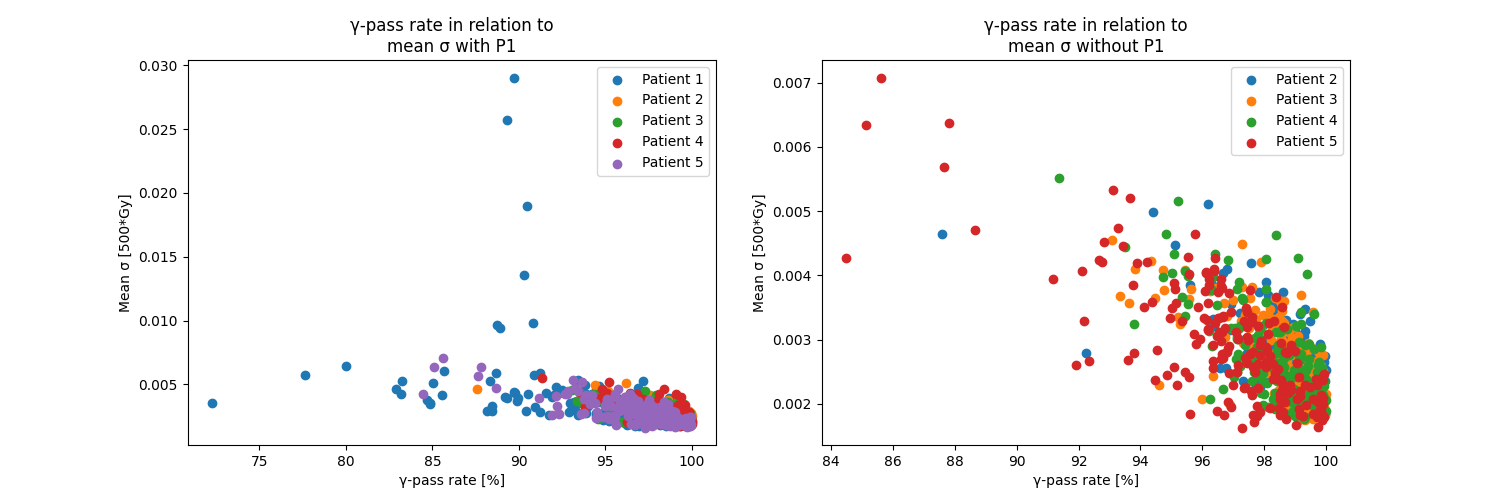}
	\caption{$[1\%, \SI{2}{\milli\m}]$ $\gamma$ pass-rate in relation to mean predicted uncertainty, e.g mean standard deviation $\sigma$. The left illustrates the results for all investigated patients while the right removes data from Patient 1.}
	\label{fig:correlations}
\end{figure*}

Correlating the prediction uncertainty on patients unseen to the model in training to the dosimetric accuracy allows us to simulate the case of dose prediction on a new patient to be irradiated, having only the predicted uncertainty as a decision making criterion. Thus, \cref{fig:correlations} depicts how the average standard deviation of dose predicted by the model correlates with the $\gamma$ pass-rate results for all patients. To avoid clustering of the predictions around \SI{100}{\percent}, a more strict criterion of $[1\%, \SI{2}{\milli\m}]$ was chosen. 

\begin{figure}[htpb]
\centering
\includegraphics[width=0.49\textwidth]{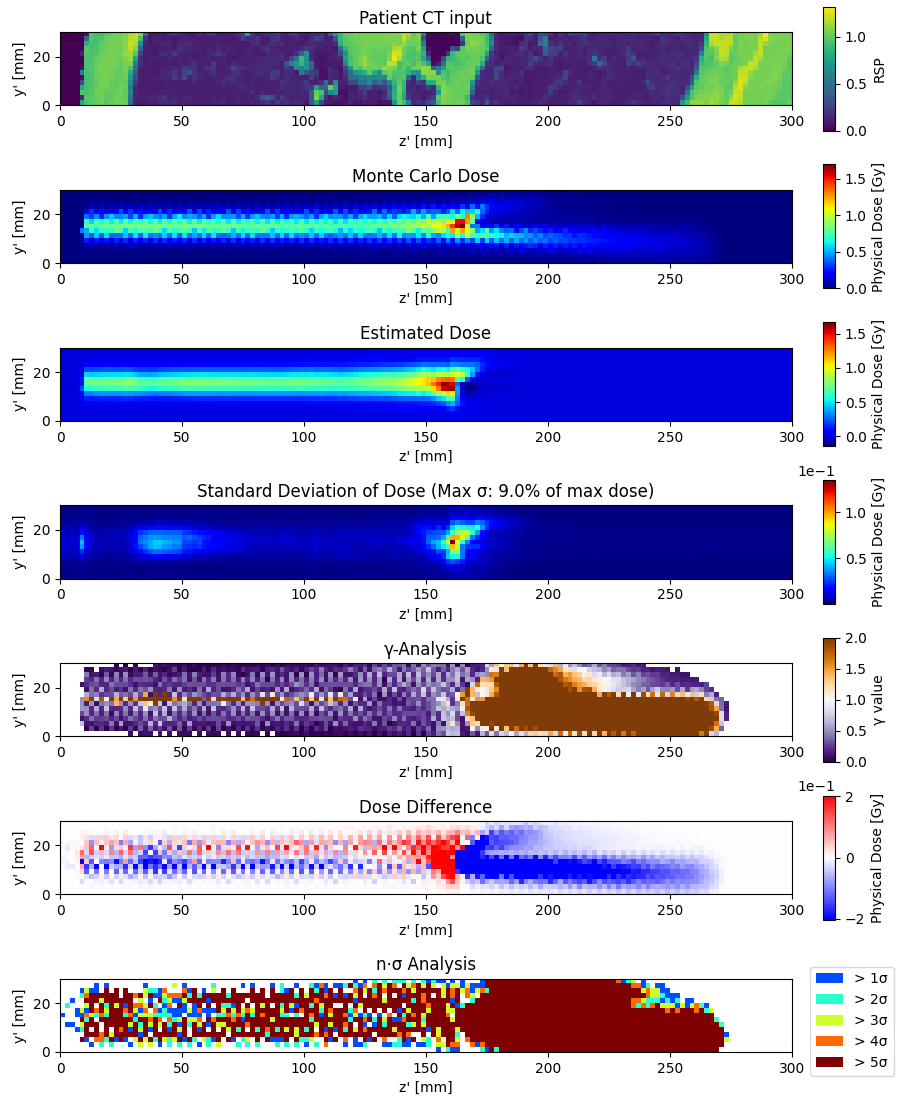}
\caption{BayesDose prediction on Patient 1 with the lowest $\gamma$ pass-rate (\SI{73.01}{\percent}) (outlier in lower left corner of \cref{fig:correlations}). The figure arrangement is similar to \cref{fig:best_phantom}.}
\label{fig:p1_worst_gamma}
\end{figure}

\begin{figure}[htpb]
\centering
\includegraphics[width=0.49\textwidth]{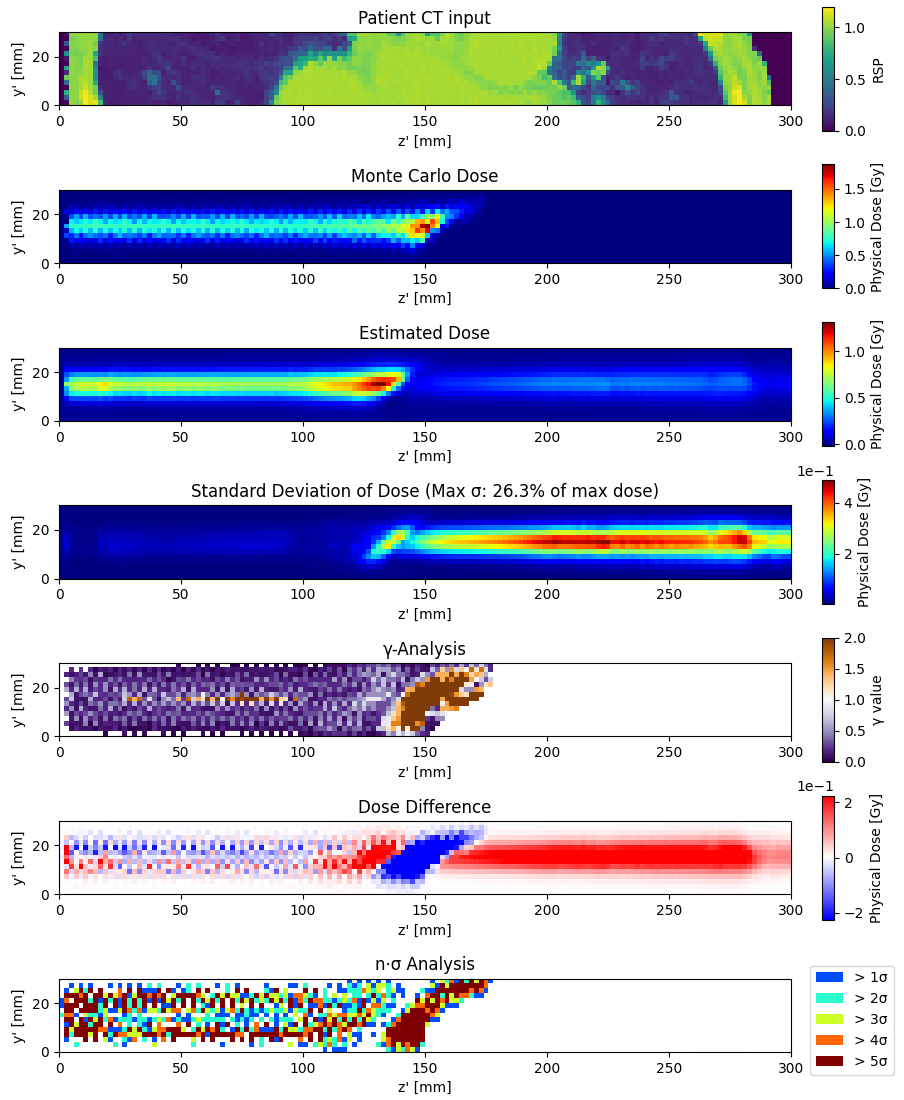}
\caption{BayesDose prediction on Patient 1 with the highest mean $\sigma$ value (outlier in upper right corner of \cref{fig:correlations}). The figure arrangement is similar to \cref{fig:best_phantom}.}
\label{fig:p1_worst_sigma}
\end{figure}

In all patients apart from Patient 1, low dosimetric accuracy correlates well with high average prediction uncertainty. Patient 1 shows sever outliers, which are investigated further below in \cref{fig:p1_worst_gamma} (worst accuracy) and \cref{fig:p1_worst_sigma} (highest average uncertainty). To quantify the correlations, we calculate correlation coefficients with and without Patient 1 in \cref{tb:correlation}. A strong negative correlation is visible, supporting how high predicted uncertainty correlates with dosimetric inaccuracies. The trend holds also among multiple $\gamma$-criteria, where strongest correlation is achieved using a $[1\%, \SI{2}{\milli\m}]$ $\gamma$ criterion (without Patient 1).

\begin{table}[htpb]
	\centering
	\caption[]{Correlation coefficients $\rho$ for different $\gamma$ pass-rates (with and without Patient 1) in correlation to mean $\sigma$}
	\begin{tabular}{l c c c c}
		\toprule
		$\gamma$ analysis & $\rho$ (All) & $\rho$ (without P1) \\
		\midrule
		$[1\%, \SI{3}{\milli\m}]$ & -0.45 & -0.64\\
		$[1\%, \SI{2}{\milli\m}]$ & -0.5 & -0.7 \\
		$[1\%, \SI{1}{\milli\m}]$ & -0.35 & -0.34 \\
		$[2\%, \SI{3}{\milli\m}]$ & -0.45 & -0.61 \\
		$[2\%, \SI{2}{\milli\m}]$ & -0.49 & -0.67 \\
		$[2\%, \SI{1}{\milli\m}]$ & -0.39 & -0.41 \\
		\bottomrule
	\end{tabular}
	\label{tb:correlation}
\end{table}

Further analysis of the outliers for Patient 1 show the following:
In the case of the worst dosimetric prediction on Patient 1, i.\,e., the far-left outlier in \cref{fig:p1_worst_gamma}, failing voxels seem to mainly originate from a low dose flair behind the Bragg peak, where the model fails to shape the distortion of the peak. 
The outlier with the highest mean standard deviation, shown in \cref{fig:p1_worst_sigma}, is also showing difficulties in predicting the dose flair distal to the Bragg peak. Further, the range seems incorrectly predicted. However, the model associates large standard deviation with the low-dose region distal to the peak, leading to high relative uncertainty. Both examples also show substantial interpolation artifacts and also highlight a common issue with using $\gamma$-tests for dosimetric analysis, particularly in this correlation study, as the low-dose region is not captured as it is discarded in the analysis by the dose threshold.

\subsection{Retraining (Experiment 5)}
To test if the substantially worse performance on Patient 1, both dosimetrically and in quality of the uncertainty prediction, can be mitigated, the model was retrained on the second worst performing dataset -- Patient 5 -- for \num{10} epochs, which also showed the second widest \ac{HU} range compared to the largest \ac{HU} range in Patient 1. \Cref{tb:transfer_learning} shows a substantial increase in dosimetric accuracy for Patient 1, similar to the other patients. 

\begin{table}[htpb]
	\centering
	\caption[]{$\gamma$-index analysis on five different lung cancer patients ($[1\%, \SI{3}{\milli\m}]$) before and after transfer learning for 10 Epochs on Patient 5.}
	\begin{tabular}{l c c c c}
		\toprule
		& Patient 1 & Patient 2 & Patient 3 & Patient 4\\
		\midrule
		Mean(\%) before TL & 96.33 & 99.29 & 99.12 & 99.32\\
		Mean(\%) after TL &   97.88 & 99.13 & 99.27 & 99.3 \\
		\bottomrule
		\toprule
	\end{tabular}
	\label{tb:transfer_learning}
\end{table}

\begin{figure}[htpb]
\centering
\includegraphics[width=0.49\textwidth]{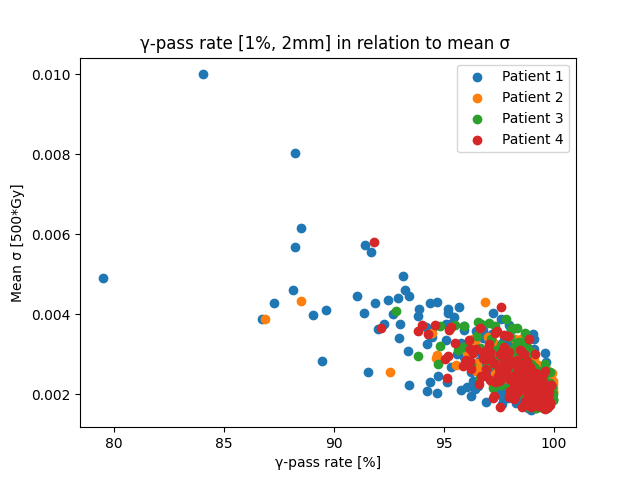}
\caption{$[\SI{1}{\percent}, \SI{2}{\milli\m}]$ $\gamma$ pass-rate in relation to mean predicted uncertainty, e.\,g., mean standard deviation $\bar{\sigma}$, after transfer learning. The correlation improves from \num{-0.5} to \num{-0.74}.}
\label{fig:new_correlation}
\end{figure}

Also, recalculating the correlation coefficients, with the newly acquired outcomes after retraining (see \cref{fig:new_correlation}), shows large improvement compared to the first evaluation in \cref{tb:correlation}, with the correlation coefficient for the illustrated $[1\%, \SI{2}{\milli\m}]$$\gamma$ pass-rate improving from \num{-0.5} to \num{-0.74}.

\subsection{Runtimes}

For a stable (average) prediction and acceptable accuracy of the standard deviation, multiple samples of the BayesDose parameters need to be taken, creating multiple predictions with the same computational operation, thus naturally increasing runtime compared to a deterministic variant of the network.

Runtimes for BayesDose computing \num{100} samples (predictions) as well as its deterministic variant and the original network are listed in \cref{tb:runtimes}. For the \enquote{naive} analysis in \cref{tb:runtimes}, the individual samples were requested fully sequential. In this naive process, a full BayesDose ensemble prediction thus took about \num{400} times as long as its deterministic variant -- a single prediction took thus about \SI{43}{\milli\second}, and the ensemble of predictions about \SI{4.3}{\second}.
However, after moving from sequential to parallel predictions reduced the runtime of the full ensemble prediction to \SI{67}{\milli\second}.
 
The deterministic variant structure and the \ac{LSTM} network from \citet{Neishabouri2021} were equally fast with a slightly faster prediction time of \SI{8}{\milli\second} for the deterministic variant of the \ac{BLSTM}. The reported measures include the time required to send the input \ac{CT} cube for each pencil beam from CPU to GPU and vice versa for the yielded dose cube.

\begin{table}[htpb]
\centering
\caption[Runtimes of trained networks for the patient dataset]{Prediction run times for the patient dataset. The reported values include the mean inference time and  \ac{std} taken by each model to predict individual beam distributions. All models were run on a Nvidia RTX A5000 GPU.}
\begin{tabular}{l S[table-format=4.1(1),separate-uncertainty=true,table-align-uncertainty = true]}
\toprule
 Model & {Runtime $[\si{\milli\second}]$} \\
\midrule
Neishabouri LSTM & 9.8+-5.4 \\
Deterministic LSTM (study) & 8+-2  \\
Bayesian LSTM naive & 4288+-565  \\
Bayesian LSTM parallelized & 67+-6.6 \\
\bottomrule
\end{tabular}
\label{tb:runtimes}
\end{table}

\section{Discussion}
In this paper, we exhibited the feasibility of using \aclp{BNN} for proton dose calculation tasks. The implemented BayesDose model features a fully Bayesian network structure built from a \ac{BLSTM} and a Bayesian backend network. It was capable of accurately predicting dose distributions inside a patient while quantifying the uncertainty in each prediction. Thus BayesDose provides supporting evidence for the viability of applying \acp{BNN} to dose calculation tasks in real-world applications. 
 
\subsection{Nominal Predictive Accuracy}
For both phantom and lung patient cases, BayesDose reproduces at least the accuracy of the original LSTM-based deterministic netowrk, both trained on an energy of \SI{104.25}{\mega\electronvolt}.  

In the phantom case study, BayesDose and its deterministic variant both showed improved metrics over the original \ac{LSTM}. 
BayesDose showed a slightly better $\gamma$ analysis performance whereas its deterministic variant had a minor improvement in \ac{MSE} and \ac{MAE} values. Note that the values presented in the results differ from the original publication on the deterministic \ac{LSTM} \cite{Neishabouri2021}, due to switching the algorithm for $\gamma$ calculation. Notable, the $\gamma$ pass-rates measured for BayesDose over the whole dataset seemed more robust than the original \ac{LSTM}. 

In the patient case, the measured test results of all three compared networks were similar with average $\gamma$ pass-rates above \SI{99.59}{\percent}.  

Training BayesDose on different initial proton energies also confirms accuracy comparable to the deterministic \ac{LSTM}-based model from \citet{Neishabouri2021}, with slight reduction in accuracy for the other energy datasets. This might originate from the higher variance in pencil beam ranges and hence a larger required clipping area to compensate for that change. However, even with a larger clipping area the number of pronounced air cavities, which led to cases where the Bragg peak was located outside the patient or reached the second lung after completely passing through the first lung, increases with higher energy ranges.

When testing the network generalization by application on other unseen patients, we also observe behavior similar to the deterministic original, with lower test results from Patient 1 and Patient 5  in comparison to the other patients. We explain these -- similarly to \citet{Neishabouri2021} -- with the different ranges of input RSP values in these patients (as shown in \cref{fig:box_plot}). Therefore, BayesDose tries to predict previously unseen RSP values, potentially failing to recognize important features such as air cavities having a drastic influence on the Bragg peak location. This hypothesis is further underlined by the results from the retraining experiment, where an increase in the \enquote{seen} \ac{RSP} range yielded substantially better results for the worst-performing Patient 1.  

These findings suggest, for both phantom and patient, that the Bayesian nature of the network structure does not show any obvious deterioration effects of prediction accuracy on the test results while being compared to an equivalent deterministic network structure. Analyzing the contributions to the loss functions during training (\cref{fig:loss_analysis}), BayesDose quickly reaches comparable performance to a deterministic model in the early epochs of model fitting. As the training progresses, BayesDose continues to explore a range of Pareto-like optimal solutions while maintaining and stabilizing test performance in terms of \ac{SMSE}. Simultaneously, it aims to optimize the probability distributions of the posterior, minimizing regularization without compromising performance. Thus, \ac{BNN} could potentially be applied to a variety of deep learning proton dose engines without major deterioration in prediction accuracy.

\subsection{Quality of the uncertainty prediction}

The defining benefit of the BayesDose over its compared deterministic variant is the quantification of uncertainty in its prediction. Visual inspection of the results from the worst dose predictions on the phantom dataset (\cref{fig:worst_phantom}) and from the patient dataset (\cref{fig:worst_patient}) both show comparably large standard deviations  -- and thus high uncertainty -- in areas where many voxels failed the $\gamma$ test as well. This spatial correlation of high uncertainty with high dosimetric differences is, while expected, a promising and also necessary feature for sensible use of the model uncertainty as a potential decision making metric.  

A more quantitative analysis was performed by measuring how dose differences between prediction and Monte Carlo relate to multiples of the standard deviation $n\sigma$. The analysis on the phantom dataset indicates an apparently \enquote{overcautious} uncertainty estimate compared to the expected probability mass when assuming a Gaussian uncertainty. A reason for this could be that wrong predictions usually occur in regions with high dose gradients, which are especially pronounced in the phantom case and whose bimodal behavior does not closely follow a Gaussian distribution, leading to large uncertainty estimates. 

In the patient dataset, the number of failing voxels move closer to what would be expected from the Gaussian assumption, but seem too low for $1\sigma$ and to large for confidence intervals greater than $3\sigma$. We attribute to the wider spectrum of anatomies as well as the interpolation artifacts, exhibiting a more \enquote{noise-like} behavior. 

\Cref{fig:best_patient} also suggests an attempt to learn the occurring interpolation artifacts as model uncertainty prediction. This is visible in form of slightly increased uncertainty around the beam line axis in the center of the dose cubes, where the interpolation effects were usually most pronounced in the training dataset. It may thus be sensible to move beyond the analysis of standard deviation in the future and, for example, examine higher moments like skewness or empirical confidence intervals quantiles.

The potential use for model quality assurance or decision making, already indicated above, is quantified by the (negative) correlation between the average $\sigma$ and the $\gamma$ pass-rate (reported in \cref{tb:correlation}). These results strongly correlate large standard deviations with inaccurate predictions. Thus, the standard deviation can potentially serve as an indicator to accept a predicted dose distribution or move to a deterministic numerical algorithm.

The strongest correlations (before re-training) could be observed at a $\gamma$ criterion of $[1\%, \SI{2}{\milli\m}]$ when excluding Patient 1. The worst performing Patient 1 features a substantially wider RSP value range which might cause the outliers from the correlation in the test results (see \cref{fig:correlations}). Most of these observed outliers, however, exhibit extreme cases of a high average standard deviation. Thus, they become neglectable for dose prediction, as cases with high uncertainty and high $\gamma$ pass-rates would be classified as false negatives.
Considering this, patient 1 only exhibits one single prediction outlier with a low $\gamma$ pass-rate and corresponding low average $\sigma$, which when used as a decision criterion would lead to a false positive (for using the BayesDose model despite limited accuracy). Further, Patient 1 could also be easily identified as not suited for model prediction during dose calculation due to its differing HU range. Hence, the model could alert the operating physician of high uncertainty in a given patient and discourage/forbid usage in this scenario.

Consequently, the BayesDose model was able to substantially increase either the $\gamma$ pass-rates results as well as the correlation coefficient by re-training on Patient 5, incorporating a larger seen \ac{RSP} value range into training. While the potential clinical use of such a model would thus involve a "preferred range of \ac{RSP}" specification, patient cases with higher \ac{RSP} values similar to patient 1 could be readily identified as unsuitable for model prediction during dose calculation due to their differing \ac{HU} range. Therefore, the model could alert the attending staff of heightened uncertainty in such patients and discourage or prohibit its usage in such scenarios. 

The accurate estimation of uncertainty in each prediction may solve the major problems of nontransparent and overconfident predictions of \acp{DNN} in areas of critical decision-making and could potentially open the doors for these algorithms in everyday clinical practice.

\subsection{Runtime}
The main limitation of \acp{BNN} is that a large ensemble of single predictions is required to have a precise prediction and estimation of uncertainty. Additionally, the network has more free parameters (depending on the probability distributions parameterizing the weights and biases), resulting in more time-consuming single predictions. This becomes apparent in \cref{tb:runtimes}, where the parallelized BayesDose network still takes about seven times as long as its deterministic variant structures for a single prediction.

However, due to the parallel nature of a neural network's feed-forward operation, there is a lot of space for additional speed improvements. By parallelizing the calculation of a large number of proton beamlets the runtime could be substantially reduced. Further, improvements can be made by developing a more efficient parallel sampling of multiple weights internally or alternatively the option to sample the required weights prior to the use case. Another speed-up could potentially be achieved by sending the \ac{CT} cubes in advance to the GPU before dose calculation, as proposed by \cite{Neishabouri2021}. This way, only the effective feed-forward time of the network would be the limiting factor in speed. 

Efforts regarding runtime optimization are further supported through constant advances in dedicated deep-learning hardware, and more prominently by leveraging the parallel nature of the problem. In future studies, the required ensemble size of predictions could also be analyzed which may lead to the conclusion that a smaller ensemble size still has satisfying precision. Thus, the computation complexity and corresponding runtime could be greatly reduced. 
Overall, the parallel structure of the \ac{BNN}'s predictions leaves multiple options for dedicated optimization and thus considerable runtime speedup during execution.

\subsection{Limitations and Application}
We demonstrated that \acp{BNN} can be used successfully for dose calculation in heterogeneous tissue. Nevertheless, in this study the implemented BayesDose model does not compose a complete dose calculation engine that is ready to be used clinically. In this study, the focus was exclusively on individual beamlets with a single initial energy (mainly of \SI{104.25}{\mega\electronvolt}) and a specific beam focus applied to a lung patient case to be able to compare the results with those of its deterministic variant proposed by \citet{Neishabouri2021}. To compose a full dose calculation engine, multiple models would need to be trained on each commissioned energy and potentially other geometric parameters. Such a dose engine for the deterministic use case is currently developed \cite{Neishabouri2023} and may afterwards be extended to use the BayesDose model presented here. While other works generalize their models by including energetic \cite{Pastor-Serrano2022} or geometric \cite[for photons]{Kontaxis2020} parameters in training, we argue that individual models exhibit no substantial disadvantage and might even be advantageous in handling extremely low and high energies in dose calculation and quality assurance.

Further, model training was limited to a single patient geometry. To generalize well over all given input images, a larger training dataset with different patient geometries as well as a variety \ac{CT} scanner should be considered for practical implementations. 

Further improvements may be achieved by using a different loss function for the model fit that has a similar dimension to the \ac{KL} divergence loss and more importantly a similar range of optimal \acp{LR}.
Hence, the efficiency of the model and its training process can be further enhanced by fine-tuning the network's parameters. 

For incorporating the measured uncertainty into the dose calculation process, a conceivable way would be to specify an uncertainty threshold (or some form of global model confidence indicator) prior to the calculation of the dose distribution. If this threshold is crossed, the algorithm could either warn about an uncertain prediction of a specific pencil beam and/or could calculate that beam distribution again with \ac{MC} simulations. Thereby, the calculation time could be substantially reduced without compromising accuracy, suggesting particular benefit to online adaptive proton therapy.

Besides the application in proton dose calculation, potential applications in photon as well as heavier ion (carbon, oxygen, helium) dose calculation would be feasible.

\subsection{Conclusion}

Our BayesDose model can achieve state-of-the-art prediction accuracy with the added benefit of comprehensively providing a spatial distribution of the model's uncertainty. 

Even with ensembles of 100 predictions, the runtime overhead stayed below a factor of 10 and could be further reduced by pruning and computational optimization. Expanding the patient training dataset to other patients and/or regular transfer learning can finetune prediction accuracy as well as uncertainty estimation. The Bayesian approach may similarly be applied to other recent and upcoming approaches in comprehensible deep-learning dose prediction. 

The strong correlation between prediction uncertainty and deviation from the ground truth could play a vital role in quality assured clinical translation of dose calculation algorithms based on neural networks.

\printbibliography

\end{document}